\documentclass[%
 reprint,
 amsmath,amssymb,
 aps,
floatfix,
]{revtex4-2}

\usepackage{graphicx}

\usepackage{dcolumn}
\usepackage{bm}
\usepackage{xcolor}
\usepackage{lineno}
\usepackage[percent]{overpic}
\usepackage{comment}
\RequirePackage[a-1b,mathxmp]{pdfx}

\begin{document}

\preprint{APS/123-QED}
\author{
D. Godagama\textsuperscript{1,2,3*}, 
M.W. Ahmed\textsuperscript{2,4},
B. Crowe\textsuperscript{2,4},
G. Feldman\textsuperscript{1},
F. Friesen\textsuperscript{5,2},
H. Gao\textsuperscript{5,2},\\
H.W. Grießhammer\textsuperscript{1},
D. Hornidge \textsuperscript{6},
C.R. Howell\textsuperscript{5,2},
H.J. Karwowski\textsuperscript{2,7},
M.A. Kovash\textsuperscript{3},
K.K.H. Leung\textsuperscript{8},\\
M. Lewis\textsuperscript{1},
C. Malone\textsuperscript{2},
E. Mancil\textsuperscript{5,2},
D. Markoff\textsuperscript{2,4},
J. A. McGovern\textsuperscript{9},
S. Mikhailov\textsuperscript{5,2},
D. R. Phillips\textsuperscript{10},
M.H. Sikora\textsuperscript{1,2},
Y.K. Wu\textsuperscript{5,2},
J. Zhou\textsuperscript{5,2}
}

\affiliation{\textsuperscript{1}Institute for Nuclear Studies, Department of Physics, The George Washington University, Washington, DC 20052, USA}
\affiliation{\textsuperscript{2}Triangle Universities Nuclear Laboratory, Durham, North Carolina 27708-0308, USA}
\affiliation{\textsuperscript{3}Department of Physics and Astronomy, University of Kentucky, Lexington, Kentucky 40506, USA}
\affiliation{\textsuperscript{4}Department of Mathematics and Physics, North Carolina Central University, Durham, North Carolina 27707, USA}
\affiliation{\textsuperscript{5}Department of Physics, Duke University, Durham, North Carolina 27708-0308, USA}
\affiliation{\textsuperscript{6}Department of Physics, Mount Allison University, 67 York St. Sackville, NB, Canada}
\affiliation{\textsuperscript{7}University of North Carolina at Chapel Hill, Chapel Hill, North Carolina 27516, USA}
\affiliation{\textsuperscript{8}Department of Physics and Astronomy, Montclair State University, Montclair, New Jersey 07043, USA}
\affiliation{\textsuperscript{9}School of Physics and Astronomy, The University of Manchester, Manchester M13 9PL, United Kingdom} 
\affiliation{\textsuperscript{10}Department of Physics and Astronomy and Institute of Nuclear and Particle Physics, Ohio University, Athens, Ohio 45701, USA}

\title{New Neutron Polarizabilities from Deuterium Compton Scattering at HI$\gamma$S}

\collaboration{COMPTON@HI$\gamma$S Collaboration}

\date{\today}

\begin{abstract}

Elastic and inelastic Compton-scattering cross sections of the deuteron were measured at HI$\gamma$S using a circularly polarized, quasi-monoenergetic $\gamma$-ray beam (61.0 MeV). Scattered $\gamma$ rays from a liquid deuterium target were detected at $55^\circ$, $115^\circ$, and $150^\circ$.\,High-resolution NaI(Tl) detectors at backward angles enabled separation of elastic and inelastic events.\,A $\chi$EFT analysis of the data yields new neutron scalar polarizabilities, $\alpha^n_{E1} = 12.25 \pm 1.1\mathrm{(stat)} \pm 0.3\mathrm{(BSR)} \pm 0.8\mathrm{(th)}$ and $\beta^n_{M1} = 2.95 \mp 1.1\mathrm{(stat)} \mp 0.3\mathrm{(BSR)} \mp 0.8\mathrm{(th)}$ (in units of $10^{-4} \mathrm{fm}^3$) reducing statistical uncertainties by 12\%.



\end{abstract}

\maketitle


Electromagnetic\,(EM)\,scalar\,polarizabilities\,($\alpha_{E1}$,\,$\beta_{M1}$) are fundamental structure constants of the nucleon. These quantities characterize the induced electric and magnetic dipole moments in response to an applied electromagnetic\,field\,\cite{GRIEHAMMER2012841}.\,Accurately\,measuring and theoretically predicting these quantities is essential for a comprehensive understanding of the nucleon's internal structure.\,Nucleon polarizabilities also provide important benchmarks for lattice QCD calculations, enabling first-principles investigations of the nucleon's electromagnetic response from its underlying quark and gluon structure \cite{Detmold2006,HAGELSTEIN201629}. Moreover, the polarizabilities play a significant role in studies of the Lamb shift in muonic hydrogen and of the proton–neutron mass difference, where they constitute the dominant source of theoretical uncertainty \cite{Pachucki,Birse2012}. Nuclear Compton scattering offers a clean probe of the polarizabilities, with the external EM field of the incident photon acting on the nucleon.

Over the past few decades, proton polarizabilities have been extracted with high accuracy through a series of Compton scattering experiments on hydrogen targets~\cite{GRIEHAMMER2012841,McGovern2013}. In contrast, neutron polarizability measurements remain limited due to inherent experimental challenges. Chief among these are the absence of dense free-neutron targets and the neutron’s lack of electric charge, which suppresses the leading-order Thomson scattering contribution.

To overcome these limitations, light nuclei such as deuterium~\cite{Lucas1994,Lundin2003,Hornidge2000,Myers2014,Strandberg2018}, $^3$He~\cite{Mancil2025,Zhou2025,zhou2026}, $^4$He~\cite{Wells1990,FUHRBERGHe41995,SikoraHe4,LiHe4} and $^6$Li~\cite{MyersLi60,MyersLi86} were used as ``neutron" targets. These nuclei provide stable and dense targets. Their coherent Compton cross sections have marked sensitivity to neutron polarizabilities, because they interfere constructively with both the proton-photon amplitude and the charged-pion exchange currents that are a consequence of the pion’s role in nuclear binding. These different targets allow for the extraction of different linear combinations of nucleon dipole polarizabilities to fully test the $\chi$EFT framework \cite{Hildebrandt2010}. Measurements of deuterium and $^4$He are sensitive to isoscalar polarizabilities ($\alpha^s_{E1}$,\,$\beta^s_{M1}$), while $^3$He is sensitive to the combination 2p+n \cite{GRIEHAMMER2012841}. The neutron polarizabilities ($\alpha^n_{E1}$,\,$\beta^n_{M1}$) can then be determined by leveraging the relatively well-known proton polarizabilities.

Compton scattering at forward angles is primarily sensitive to the sum of the scalar polarizabilities, $\alpha + \beta$, while backward-angle scattering cross sections are more sensitive to the difference, $\alpha - \beta$ \cite{Wissman2004,GRIEHAMMER2012841}. However, $\alpha + \beta$  is independently constrained by the Baldin Sum Rule (BSR) \cite{BALDIN1960}, which relates it to the total photoabsorption cross section. Thus, backward-angle data are sufficient to determine individual polarizabilities, while forward data provide a consistency check with the BSR. As new nucleon photoabsorption data became available, the BSR values for the proton and neutron were re-evaluated. The most recent values are (given in the units of $10^{-4}\ \mathrm{fm^3}$ that are used hereafter) \cite{PhysRevD.92.074031, Levchuk2000}.
\begin{equation}
\label{Eqn:1}
\begin{aligned}
& \alpha^{p}_{E1}+\beta^{p}_{M1} = 14.0\pm0.2  \\
& \alpha^{n}_{E1}+\beta^{n}_{M1} = 15.2\pm0.5 \\
\therefore\ & \alpha^{s}_{E1}+\beta^{s}_{M1} = 14.6\pm0.3
\end{aligned}
\end{equation}

The current world data set for deuterium Compton-scattering measurements comprises only 50 data points and 2 outliers 
with photon energies spanning 49-112 MeV \cite{Lucas1994,Lundin2003,Hornidge2000,Myers2014}. It is smaller, of lower quality, and spans a narrower energy range than the corresponding proton data which contain 220 data points up to 170 MeV and features unpolarized and linearly polarized cross sections \cite{McGovern2013,Li,Mornacchi}. The BSR-constrained $\chi$EFT analysis of the current deuteron elastic Compton-scattering data yields the following values for the isoscalar-averaged scalar polarizabilities with $\chi^2$/dof = 45.2/44 = 1.03 \cite{Myers2014}.
\begin{equation}
\begin{aligned}
\alpha^s_{E1} &= 11.1 \pm 0.6\,\mathrm{(stat)} \pm 0.2\,\mathrm{(BSR)} \pm 0.8\,\mathrm{(th)} \\
\beta^s_{M1}  &=  3.4 \mp 0.6\,\mathrm{(stat)} \pm 0.2\,\mathrm{(BSR)} \mp 0.8\,\mathrm{(th)}
\end{aligned}
\end{equation}
Combining the above with the proton polarizabilities \cite{McGovern2013} leads to
\begin{equation}
\begin{aligned}
\alpha^n_{E1} &= 11.55 \pm 1.25\,\mathrm{(stat)} \pm 0.2\,\mathrm{(BSR)} \pm 0.8\,\mathrm{(th)}, \\
\beta^n_{M1}  &=  3.65 \mp 1.25\,\mathrm{(stat)} \pm 0.2\,\mathrm{(BSR)} \mp 0.8\,\mathrm{(th)}.
\end{aligned}
\end{equation}
In deuteron Compton scattering, an inelastic contribution to the cross section arises from the breakup of the deuteron by the incident photon. Owing to the deuteron’s low binding energy of 2.2 MeV, this breakup contribution lies very close to the elastic scattering peak. The inelastic component becomes particularly prominent at backward scattering angles, where the momentum transfer between the photon and the nucleus is larger.\,Therefore, this contribution must be identified and separated to ensure an accurate extraction of the nucleon polarizabilities.\,Resolving such structure requires high energy resolution in both the incident photon beam and the photon detectors and also some theoretical knowledge of the inelastic energy distribution. Recently, the $\chi$EFT formalism has been extended to include inelastic scattering processes, specifically the deuteron breakup below 100 MeV \cite{Liao}. Accordingly, the present experiment was designed not only to extract elastic cross sections but also to measure inelastic cross sections with comparable accuracy, thereby providing an experimental benchmark for these recent theoretical calculations.



The experiment was conducted at the High Intensity $\gamma$-ray Source (HI$\gamma$S) facility at the Triangle Universities Nuclear Laboratory (TUNL). HI$\gamma$S utilizes an ultraviolet free-electron laser (FEL) and Compton backscattering to generate a quasi-monoenergetic $\gamma$-ray beam with linear or circular polarization \cite{FEL}. The energy spread of the $\gamma$-ray beam is defined by a passive collimator positioned 52.8\,m downstream of the electron-photon collision point located at the center of the FEL cavity. In this experiment, a circularly polarized $\gamma$-ray beam was used with an 8\,mm diameter collimator, resulting in a beam energy spread of 2.1\% (1.3 MeV) full width at half maximum (FWHM) and an intensity-weighted average energy of 61.0\ MeV.

The $\gamma$-ray beam was incident on a liquid deuterium target held in a conical frustum cell 20\,cm in length and 4\,cm in diameter \cite{KENDELLEN2016174}.\,The cell was constructed from a 0.125\,mm-thick Kapton foil.\,It was housed inside an aluminum vacuum can with entrance and exit windows made from the same Kapton material. The deuterium inside the cell was maintained at 19.0\,K, using a cryocooler-based system yielding a target thickness of $(9.68 \pm 0.13) \times 10^{23}$ nuclei/cm$^2$. The target was periodically emptied to collect data (empty-target data) to subtract the background induced by the Kapton windows.

The photon flux was monitored using a thin plastic scintillator paddle placed near the downstream mirror of the FEL cavity. The paddle's efficiency was measured using a NaI(Tl) detector located downstream of the target. Flux to the in-beam NaI detector was attenuated using a series of copper attenuators to limit the counting rate of the detector. The measured flux was corrected to account for the attenuation through copper. The photon flux on target during the experiment was $(1.000\pm0.033)\times10^7$\,$\gamma$/s.

\begin{figure}[b]
\centering
\includegraphics[width=\linewidth]{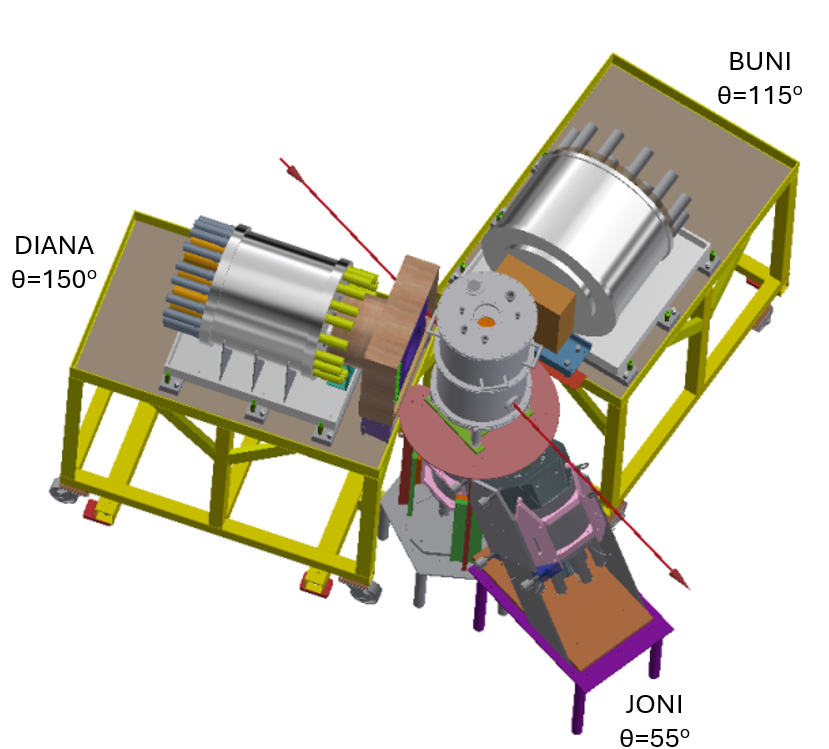}
\caption{\label{fig:ExpSetup} 3D rendering of the experimental setup showing the layout of the cryogenic target and the NaI(Tl) detectors. The direction of the incident photons is indicated by the red arrow.}
\end{figure}

The scattered $\gamma$ rays from the target were detected by three NaI(Tl) detectors placed at laboratory scattering angles $55^\circ$, $115^\circ$, and $150^\circ$ named JONI, BUNI and DIANA, respectively (shown in Fig.\ \ref{fig:ExpSetup}). 

DIANA and BUNI \cite{BUNINIM} are large-volume detectors with energy resolutions ($\Delta E/E$) of 2.0\% and 2.4\%, respectively, at 61 MeV. JONI is smaller, and has an energy resolution of approximately 10\% at the same energy. DIANA consists of a cylindrical single-crystal NaI(Tl) core measuring 50.8 cm in length and 48.3 cm in diameter, surrounded by a 4.0 cm-thick segmented NaI(Tl) annulus. BUNI has a cylindrical NaI(Tl) core (26.7 cm diameter, 55.9 cm length) encircled by an 11.4 cm-thick NaI(Tl) annulus segmented into four quadrants. The quadrants are individually calibrated and summed with the core response. Its NaI(Tl) segments are further surrounded by a 12.7\,cm-thick BC400 plastic scintillator annulus segmented into six sections. JONI's NaI(Tl) core (27.9 cm length and 25.4 cm diameter) is surrounded by a 7.6 cm-thick NaI(Tl) annulus segmented into eight optically isolated sections. All annuli are used for cosmic-ray rejection, while seven 6\,mm-thick BC600 plastic scintillator sheets placed around the top and sides of DIANA further enhance cosmic-ray rejection. All detectors have a $\sim$ 15 cm thick lead collimator installed at the detector front, with a geometry optimized to allow the photons scattered from the target to fully illuminate the back face of the core. 

Data were recorded using 14-bit CAEN V1730 waveform digitizers with a 2 $\mu$s digitization window. Data from the digitizer’s internal memory were read using a STRUCK SIS3153 controller module via the VME backplane and transferred to the data acquisition computer \cite{Godagama2022}. In addition to signals from the detectors, the period of the electron storage ring radio-frequency (RF) signal of 358\,ns was also digitized to serve as a timing reference. The time-of-flight (ToF) for each event was defined as the time interval between the event trigger and the next rising edge of the RF signal. Digitized waveforms were processed with baseline subtraction, and energy depositions were extracted via pulse integration. Timing information was obtained using a software-based constant fraction discrimination (CFD) algorithm applied to the digitized pulses.

Before positioning the detectors at their scattering angles, both DIANA and BUNI were placed directly facing the $\gamma$-ray beam to record the energy spectrum of the incident beam (referred to as the in-beam data). These were used to determine the beam energy distribution.

A geometrical survey of the experimental setup was performed, and the results were incorporated into a Monte Carlo (MC) simulation using the GEANT4 toolkit \cite{GEANT4,GEANT4:2}. This simulation was used to extract effective solid angles which accounted for the geometric effects due to the extended target, finite acceptance of the detectors, attenuation of the incident photons in target material and Kapton windows, and of the scattered photons in the target and surrounding materials. The simulated effective solid angles for detectors JONI, BUNI and DIANA were 21.4, 46.5 and 63.2 msr, respectively.

\begin{figure}[h]
\includegraphics[width=\linewidth]{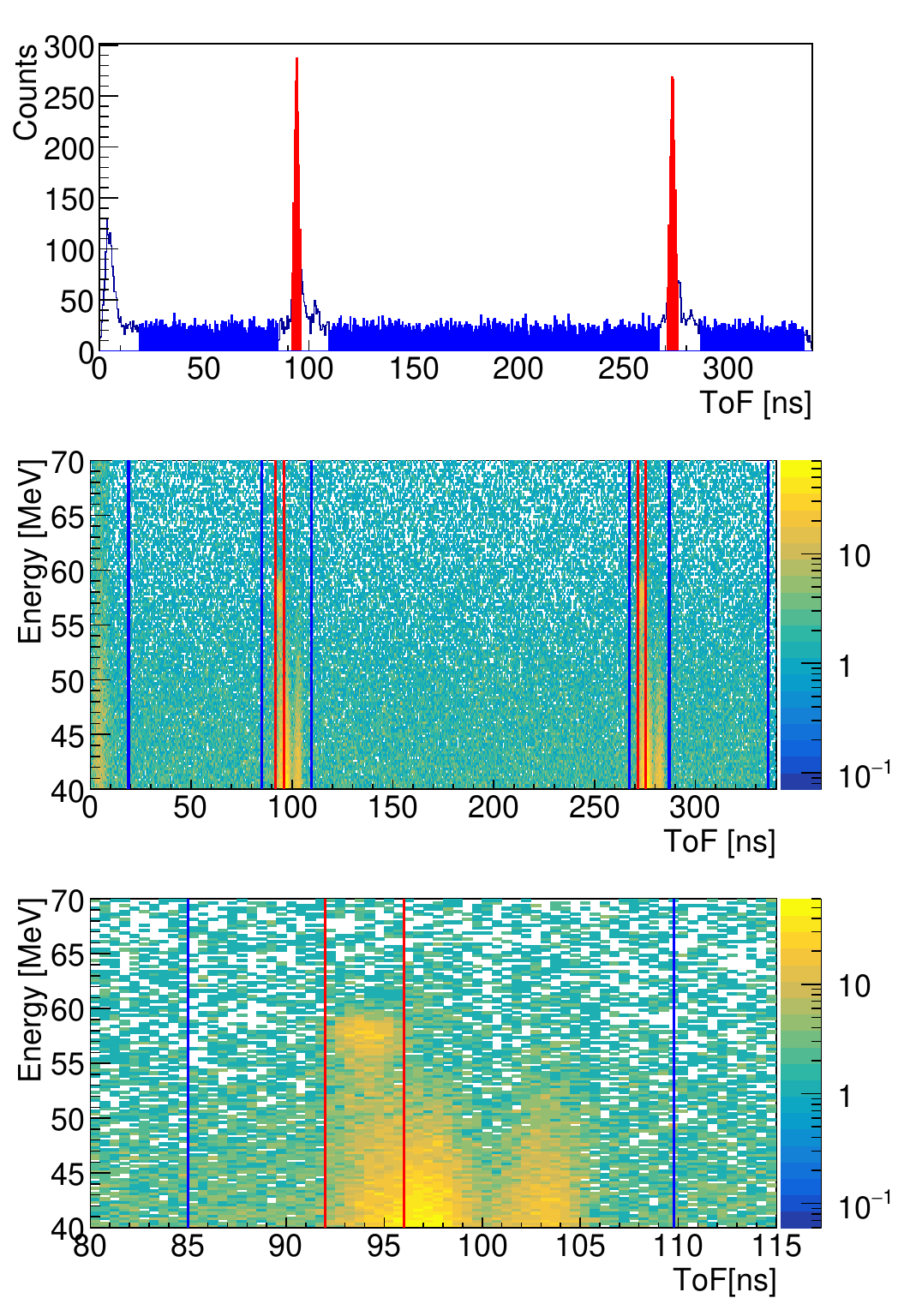}
\caption{\label{fig:BUNI_ToFCut} (Top) BUNI’s ToF spectrum subjected to the energy cut of 55.0--62.0 MeV. The counts due to the prompt $\gamma$-ray pulse on target are shaded in red while counts due to the events detected at random times relative to the beam pulse are shaded in blue. (Middle) 2-D spectrum of BUNI’s energy deposition vs ToF. Prompt and random timing windows are indicated with vertical red and blue lines respectively. (Bottom) The same spectrum shown in the middle with a zoomed-in view of the first prompt region.}
\end{figure}

The dominant source of background in this experiment was cosmic-ray muons, which were suppressed using two complementary techniques. First, cosmic muons deposit significantly more energy in the shield segments of the detectors than the electromagnetic shower leakage from Compton-scattered $\gamma$-rays. This difference enabled the application of energy-based veto cuts on the annular segments and plastic scintillators to reject cosmic events. Second, the timing structure of the $\gamma$-ray beam produced a distinct prompt peak sitting on a flat background of beam-uncorrelated events in the ToF spectrum (shown in Fig.\ \ref{fig:BUNI_ToFCut}). Beam-induced events were selected by placing a time cut around the prompt ToF peak.

\begin{figure*}[t]
\includegraphics[width=\textwidth]{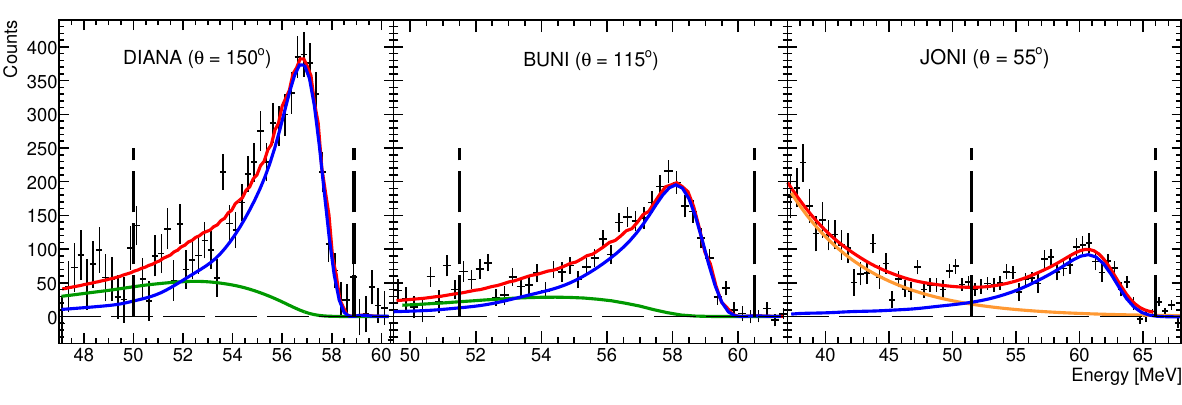}
\caption{\label{fig:lineShapFit} The final energy spectra. The curves shown are elastic line-shape (blue), inelastic line-shape (green), low energy electromagnetic background (orange) and the total fit (red). Vertical dashed lines indicate the
ROI.}

\end{figure*}

The combined annulus energy and prompt ToF window cuts suppressed the cosmic background by more than 99\%. The residual time-uncorrelated background was further reduced by sampling the energy spectrum corresponding to the events within a random timing region of the ToF distribution and subtracting it from the prompt region, with appropriate normalization to the relative widths of the timing windows. The prompt and random timing windows for BUNI are shown in Fig.\ \ref{fig:BUNI_ToFCut}.

This analysis was carried out separately for both the full-target and empty-target data. Finally, the energy spectrum from the empty-target data was subtracted from the full-target energy spectrum with proper normalization to the relative incident photon flux to obtain the net-yield energy spectra.



The net-yield spectra were then fitted using detector line-shape functions derived using a combination of MC simulations and Fourier transformation methods, as detailed below.\,For DIANA and BUNI, the line-shape includes two components to represent elastic Compton scattering and inelastic scattering. In contrast, only a broad elastic line-shape was used for JONI.

The in-beam energy spectra measured with DIANA and BUNI represent the convolution of the incident photon-beam energy distribution with the detector response. The intrinsic energy resolutions of both detectors have been well characterized using tagged photons at MAX-lab \cite{Khayrullo}. To extract the incident beam-energy distribution from the measured spectra, the detector response must first be characterized using MC simulations and the known intrinsic energy resolution.

For this purpose, a 61.0 MeV monochromatic $\gamma$-ray beam was directed onto the front face of each detector in the MC simulation, and the energy deposited in the detector was histogrammed. The resulting spectrum was then smeared with a Gaussian function whose standard deviation was calculated as a function of $\sqrt{E}$ to account for the known intrinsic detector resolution. The resulting detector response function was subsequently used to deconvolute the incident beam-energy distribution from the measured in-beam spectrum using the Gold deconvolution algorithm \cite{GOLDDcon,SUN2009312}.

With the incident beam-energy distribution thus determined, a second MC simulation was performed to obtain the expected elastic line-shape under the experimental geometry. In this simulation, each detector was placed at its scattering angle, and monochromatic 61.0 MeV $\gamma$ rays were emitted from a source representing the intersection volume of the incident beam and the liquid deuterium target, directed into the detector's solid angle with the Compton kinematic energy shifts applied. The resulting spectrum was then convoluted with the extracted incident beam-energy distribution using the Fourier transformation to obtain the elastic line-shape.

The inelastic line-shape was constructed using the above-mentioned theoretical calculation of the inelastic channel in $\chi$EFT. The detector response to inelastically scattered photons was modeled by a MC simulation in which photons with energies sampled from the theoretical distribution were emitted into the solid angle of the detectors from a source representing the intersection volume of the $\gamma$-ray beam and the liquid deuterium target. The detectors were placed at experiment scattering angles and the energy deposition in the detectors was histogrammed. The same Gaussian smearing mentioned above was applied to the spectrum to account for the intrinsic detector resolution. The resulting spectrum was then convoluted with the beam energy distribution to obtain the inelastic line-shape. Thus, the ``shape" of the inelastic line-shape is determined by the $\chi$EFT calculation, while its magnitude is obtained by fitting to the data.  
Consequently, the elastic and inelastic line-shapes were simultaneously fitted to the measured energy spectrum, with their amplitudes treated as independent fit parameters. The separation between the elastic and inelastic line-shapes was fixed, while the amplitude of each component was varied to minimize the $\chi^2$ of the overall fit. The resulting fits to the DIANA and BUNI energy spectra, shown in the left and middle panels of Fig.~\ref{fig:lineShapFit}, yielded $\chi^2/\mathrm{dof}$ values of 1.35 and 1.47, respectively.

The fit to JONI spectrum is shown in Fig.\ \ref{fig:lineShapFit} (right) with $\chi^2/\mathrm{dof} = 1.42$. The prominent background visible in the low-energy region of the JONI energy spectrum is due to atomic Compton scattering and was assumed to have an exponential shape. Therefore, a function of the following form was used to fit this background,
$f (E) = e^{-p_{0}*E} + p_{1}$ where $p_{0}$ and $p_{1}$ are the fitting parameters. This function was simultaneously fitted with the line-shape of the JONI detector. Unlike in the cases of DIANA and BUNI, JONI’s intrinsic energy resolution is not well characterized. Therefore, the smearing was done by convoluting the line-shape with a Gaussian function of free parameters. These parameters were also allowed to vary, minimizing the $\chi^{2}$ of the total fit to the energy spectrum.

The net total yields were obtained by integrating the energy spectra within the region of interest (ROI). The net elastic and inelastic yields were extracted by scaling the total yield with the ratio of the integral of the elastic (or inelastic) line-shape to the total fit within the ROI. The yields were then normalized to the total number of incident photons, target thickness and effective solid angles to calculate the differential cross sections. 

\begin{table}[htbp]
\caption{\label{tab:table1}%
The measured elastic and inelastic Compton scattering cross sections of deuterium
}
\begin{ruledtabular}
\begin{tabular}{ccccc}
\ & $\theta_{Lab}$ & $d\sigma$/$d\Omega$ & Tot. Stat. & Tot. Syst. \\
Elastic/Inelastic & & (nb/sr) & (nb/sr)& (nb/sr) \\
\colrule
Elastic+Inelastic & 55$^{\circ}$ & 15.8 & $\pm0.5$ & $\pm0.6$\\
Elastic & 115$^{\circ}$ & 13.2 & $\pm0.4$ & $\pm0.5$\\
Elastic & 150$^{\circ}$ & 17.0 & $\pm0.6$ & $\pm0.7$\\   
Inelastic & 115$^{\circ}$ & 2.6 & $\pm0.2$ & $\pm0.4$\\
Inelastic & 150$^{\circ}$ & 3.4 & $\pm0.2$ & $\pm0.5$\\
\end{tabular}
\end{ruledtabular}
\end{table}

\begin{figure}[h]
\includegraphics[width=\linewidth]{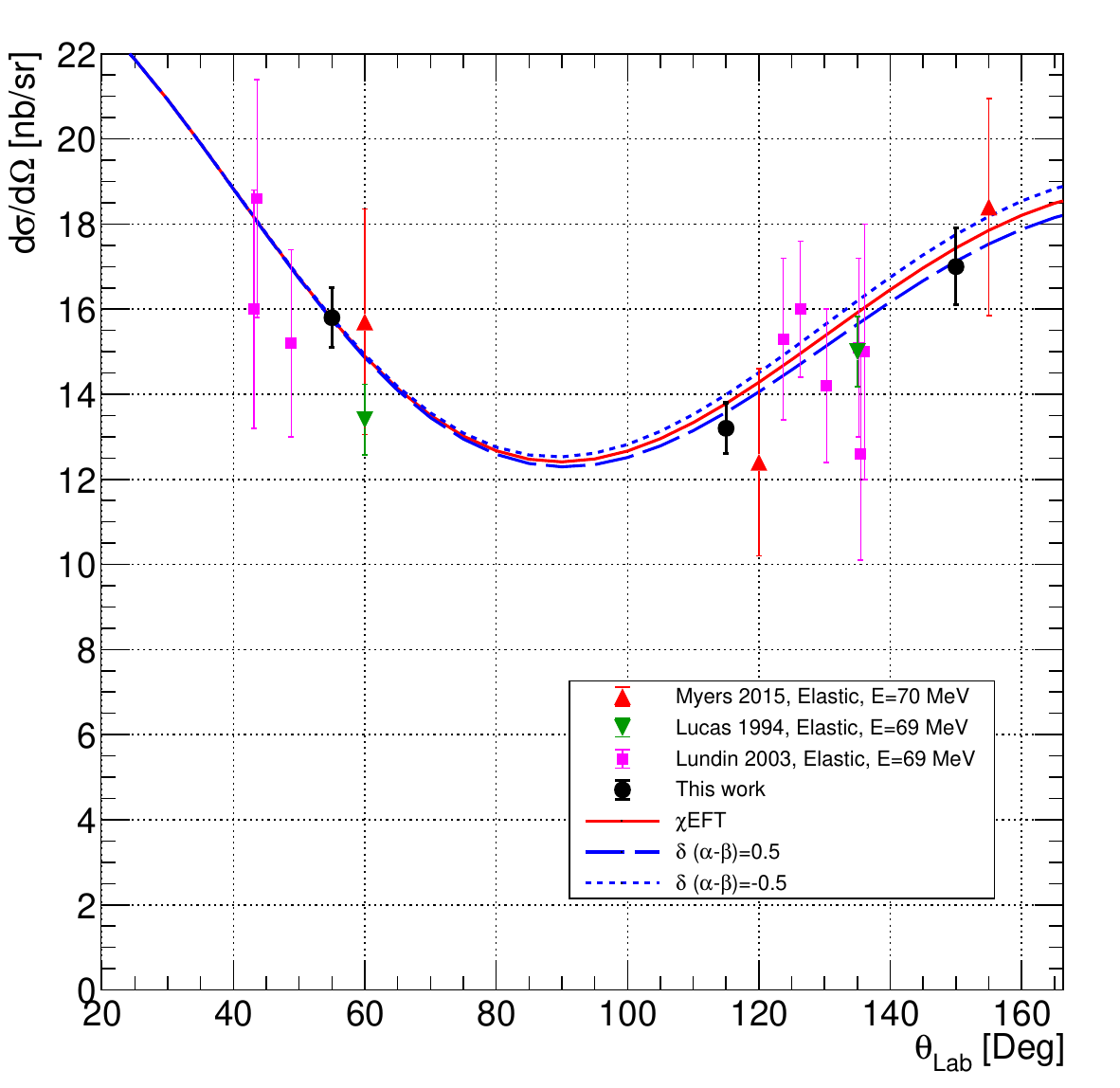}
\caption{\label{fig:AllXs} Laboratory elastic Compton scattering cross sections measured by this work compared with the results from previous experiments. 
Red curve: $\chi$EFT fit of new world data, eq.~\eqref{eq:newworld} with statistical uncertainties.}
\end{figure}

The measured elastic and inelastic differential cross sections are given in Table \ref{tab:table1}, and Fig.\ \ref{fig:AllXs} shows the angular distribution of the elastic cross sections from this work, compared to previous measurements. The red curve in Fig.\ \ref{fig:AllXs} indicates the $\chi$EFT fit subject to the BSR of eq.~\eqref{Eqn:1} to the new world data with the results from this experiment. The Myers \cite{Myers2014}, Lucas \cite{Lucas1994}, and Lundin \cite{Lundin2003} measurements shown in Fig.\ \ref{fig:AllXs} were taken at slightly different photon energies, but the plot shows no energy correction since its impact is minimal. At 55$^\circ$ the elastic and inelastic contributions could not be reliably separated and are therefore reported as their combined cross section. 


The $\chi$EFT calculation \cite{Liao} predicts inelastic cross sections of 1.9$\pm$0.2 and 2.6$\pm$0.3 nb/sr at $115^\circ$ and $150^\circ$, respectively. The measured inelastic cross sections are in reasonable agreement with these predictions within the combined experimental and theoretical uncertainties.
The systematic uncertainties in our analysis are summarized in Table \ref{tab:table3} and are categorized into two types. The first category, normalization uncertainties, affects all data points uniformly and includes contributions from the measurement of the incident-photon flux and the target thickness. The second category, point-to-point uncertainties, varies between data measured with different detectors and arises from the placement of cosmic-ray shield cuts, ToF cuts, and the selection of the ROI in the energy spectra used to extract yields. The uncertainty associated with the solid angle calculation was negligible due to the high precision of the geometrical survey of the experimental setup. All systematic uncertainties were summed in quadrature to obtain the total systematic uncertainty.

\begin{table}[h]
\caption{\label{tab:table3}%
Sources of systematic uncertainties 
}
\begin{ruledtabular}
\begin{tabular}{ccc}
\textrm{Type}&
\textrm{Source}&
\textrm{Value}\\
\colrule
Normalization & Photon flux & 3.3\% \\
 & Target thickness & 1.3\% \\ 
Point-to-point& ToF cuts at $55^\circ,115^\circ$& 0.7\% \\
    &   ToF cuts at $150^\circ$& 2.0\%\\
    & Shield cuts & 0.1\% \\
    & Elastic fit \& ROI & 0.5\% \\
    & Inelastic amp. fit parameter ($115^\circ$) & 12.6\% \\ 
    & Inelastic amp. fit parameter ($150^\circ$) & 12.2\% \\ 
\end{tabular}
\end{ruledtabular}
\end{table}

Owing to the high photon flux at HI$\gamma$S, improved beam and detector energy resolution, and enhanced statistics at backward angles, we report the highest accuracy deuteron Compton-scattering cross sections to date.

The isoscalar polarizabilities were extracted by fitting our results both independently and in conjunction with the world data. The elastic-scattering cross sections were fitted using the $\chi$EFT calculation of \cite{GRIEHAMMER2012841}, with $\alpha_s$, $\beta_s$, and an overall normalization factor as free parameters. The $\chi$EFT fit to our data without the BSR constraint applied results in

\begin{equation}
\begin{aligned}
\alpha^s_{E1} &= \  12.6 \pm\ 2.3\,\mathrm{(stat)}, \\
\beta^s_{M1} &= \ 1.8 \pm\ 1.3\,\mathrm{(stat)}.
\end{aligned}
\end{equation}
Therefore, $\alpha^s_{E1}+\beta^s_{M1}=14.4 \pm 2.6$. This sum is consistent with the BSR (\ref{Eqn:1}) and hence justifies using the BSR to constrain the fit and reduce the statistical uncertainty.

The BSR-constrained $\chi$EFT fit to our data results in 


\begin{equation}
\alpha^s_{E1}-\beta^s_{M1} = 10.9\pm2.3\,\mathrm{(stat)}\pm0.8\,\mathrm{(th)}
\end{equation}

with $\chi^2$/dof = 0.21/1 = 0.21 which then leads to

\begin{equation}
\begin{aligned}
\alpha^s_{E1} &= 12.7 \pm 1.2\,\mathrm{(stat)} \pm 0.2\,\mathrm{(BSR)} \pm 0.8\,\mathrm{(th)}, \\
\beta^s_{M1}  &=  1.9 \mp 1.2\,\mathrm{(stat)} \pm 0.2\,\mathrm{(BSR)} \mp 0.8\,\mathrm{(th)}.
\end{aligned}
\end{equation}
When our data are incorporated into the world data set, the BSR-constrained $\chi$EFT fit yields
\begin{equation}
\alpha^s_{E1}-\beta^s_{M1} = 8.4\pm1.0\,\mathrm{(stat)}\pm0.8\,\mathrm{(th)}    
\end{equation}
with $\chi^2$/dof = 46.8/46 = 1.02. This results in
\begin{equation}
\label{eq:newworld}
\begin{aligned}
\alpha^s_{E1} &= 11.5 \pm 0.5\,\mathrm{(stat)} \pm 0.2\,\mathrm{(BSR)} \pm 0.8\,\mathrm{(th)}, \\
\beta^s_{M1}  &=  3.1 \mp 0.5\,\mathrm{(stat)} \pm 0.2\,\mathrm{(BSR)} \mp 0.8\,\mathrm{(th)}.
\end{aligned}
\end{equation}

We find that the inclusion of our data in the world data set reduces the current statistical uncertainties of $\alpha^s_{E1},\beta^s_{M1}$ by 16.7\%. This is a markedly larger improvement than would typically be expected from the addition of only three data points. This is attributed to the substantially smaller statistical and systematic uncertainties of our data compared to previous measurements. These results underscore the value of high-precision data in constraining the extraction.
Combined with the proton analysis of refs.~\cite{McGovern2013, GRIEHAMMER2012841, Griesshammer:2015ahu, Li}, we therefore find for the neutron
\begin{equation}
\begin{aligned}
\alpha^n_{E1} &= 12.25 \pm 1.1\,\mathrm{(stat)} \pm 0.3\,\mathrm{(BSR)} \pm 0.8\,\mathrm{(th)}, \\
\beta^n_{M1}  &=  2.95 \mp 1.1\,\mathrm{(stat)} \mp 0.3\,\mathrm{(BSR)} \mp 0.8\,\mathrm{(th)},
\end{aligned}
\end{equation}
which is statistically consistent with the previous extraction~\cite{Myers2014} but reduces the uncertainty by $12\%$. An upcoming publication will provide details of the new determination of isoscalar and neutron values~\cite{newfitdetails}.

In summary, we measured deuteron Compton scattering cross sections with the highest accuracy achieved to date using the intense, circularly polarized, quasi-monoenergetic photon beam at HI$\gamma$S, at an intensity-weighted average photon energy of 61.0 MeV.\,At backward scattering angles, the elastic and inelastic scattering processes were resolved, allowing their cross sections to be measured separately. The elastic scattering data were analyzed within the $\chi$EFT framework to extract the isoscalar polarizabilities of the nucleon. Inclusion of our data into the current world database of deuterium Compton measurements significantly improved the accuracy of the $\chi$EFT extraction of the isoscalar polarizabilities, reducing the associated uncertainty by 16.7\%. Combining the extracted isoscalar polarizabilities with the current proton polarizabilities yielded new neutron scalar polarizabilities, with a 12\% reduction in their statistical uncertainties. Inelastic cross sections provided first benchmark to the newly available $\chi$EFT calculation of the inelastic channel. These results demonstrate the impact of high-precision Compton-scattering measurements on constraining the polarizabilities.


\begin{acknowledgments}
This work is funded in part by the U.S. Department
of Energy under grants No. DE-SC0016656 (MAK), DE-SC0016581 (GF), DE-FG02-97ER41033 (CRH), DE-FG02-03ER41231 (HG), DE-FG02-97ER41041 (HJK), DE-SC0005367 (MWA), DE-SC0025227 (MWA), DE-SC0015393 (HWG) and DE-FG02-93ER-40756 (DP) as well as by the UK Science and Technology Facilities Council award ST/V001116/1. We acknowledge the support of the HI$\gamma$S accelerator scientists, engineers, and operators for assisting with the experimental setup and for the high-quality production of the $\gamma$-ray beam. We also thank J.~Liao for providing results for the inelastic channel. 
\end{acknowledgments}

\nocite{*}
\bibliography{apssamp}

\end{document}